\documentclass[12pt,preprint]{aastex}
\renewcommand{\cite}{\citealp}

\newcommand{\realfigure}[3]{ 
             \hbox{~} \centerline{\includegraphics[width=3.2in]{#1}}
             \figcaption{#2 \label{#3}} \vspace{0.05in}\centerline{}}

\shorttitle{Variable stars in the CVn~ I dSph}
\shortauthors{Kuehn et al.}

\begin{document}

\title{Variable Stars in the Newly Discovered Milky Way Dwarf Spheroidal
Satellite Canes 
Venatici~~I\altaffilmark{1}}
\author{Charles Kuehn\altaffilmark{2}, Karen Kinemuchi\altaffilmark{3,4},
Vincenzo Ripepi\altaffilmark{5}, Gisella Clementini\altaffilmark{6}, Massimo 
Dall'Ora\altaffilmark{5}, Luca Di Fabrizio\altaffilmark{7},Christopher T. 
Rodgers\altaffilmark{3}, Claudia Greco\altaffilmark{6}, Marcella 
Marconi\altaffilmark{5}, Ilaria Musella\altaffilmark{5}, Horace A. 
Smith\altaffilmark{2}, M\'arcio 
Catelan\altaffilmark{8}, Timothy C. Beers\altaffilmark{2,9}, and Barton J. 
Pritzl\altaffilmark{10,11}}
\altaffiltext{2}{Department of Physics and Astronomy, Michigan State
University, East 
Lansing, MI 48824, USA; kuehncha@msu.edu, smith@pa.msu.edu, beers@pa.msu.edu}
\altaffiltext{3}{University of Wyoming, Department of Physics and Astronomy,
Laramie, WY 82071, USA; crodgers@uwyo.edu}
\altaffiltext{4}{Current address: Universidad de Concepcion, Departamento de
Fisica, Concepcion, Chile, and University of Florida, Department of
Astronomy, Gainesville, FL 32611-2055, USA;
kinemuchi@astro.ufl.edu}
\altaffiltext{5}{INAF, Osservatorio Astronomico di Capodimonte, Napoli, Italy; 
dallora@na.astro.it,ripepi@na.astro.it, marcella@na.astro.it, 
ilaria@na.astro.it}
\altaffiltext{7}{INAF, Centro Galileo Galilei and Telescopio Nazionale Galileo,
Santa Cruz de La Palma, Spain; difabrizio@tng.iac.es}
\altaffiltext{6}{INAF, Osservatorio Astronomico di Bologna, Bologna, 
Italy; gisella.clementini@oabo.inaf.it, claudia.greco@oabo.inaf.it}
\altaffiltext{8}{Pontificia Universidad Cat$\rm{\acute{o}}$lica de Chile,
Departamento de 
Astronom\'{\i}a y Astrof\'{\i}sica, Santiago, Chile; mcatelan@astro.puc.cl}
\altaffiltext{9}{Joint Institute for Nuclear Astrophysics, Michigan State
University, East 
Lansing, MI 48824, USA}
\altaffiltext{10}{Macalester College, Saint Paul, MN 55105, USA}
\altaffiltext{11}{Current address: Department of Physics and Astronomy, 
University of Wisconsin Oshkosh, Oshkosh, WI 54901, USA; pritzlb@uwosh.edu}

\altaffiltext{1}{Based on data collected at the 2.3m
telescope at the Wyoming Infrared Observatory (WIRO) at Mt. Jelm, Wyoming, USA, 
and 
at the INAF-Telescopio Nazionale Galileo and the 4.2m William Herschel 
Telescope, at 
Roche de los Muchachos, Canary Islands, Spain.}

\begin{abstract}
We have identified 23 RR Lyrae stars and 3 possible Anomalous Cepheids among 84 
candidate variables in the recently discovered 
Canes Venatici I dwarf spheroidal galaxy.  The mean period of 18 RRab type stars 
is 
$\langle P_{ab}\rangle = 0.60 \pm0.01$ days.
This period, and the location of these stars in the period-amplitude 
diagram, suggest that Canes Venatici I is likely an Oosterhoff-intermediate 
system.  The average apparent 
magnitude of the RR Lyrae stars,  ${\rm \langle V\rangle} = 22.17 \pm 0.02$ mag,
is used to obtain a precision distance estimate of 210$^{+7}_{-5}$ kpc, for an 
adopted reddening $E(B-V)$=0.03 mag.  We present a $B,V$ 
color-magnitude 
diagram of Canes Venatici~I that reaches $V \sim$ 25 mag,  and shows that the 
galaxy has a mainly old stellar population with a metal 
abundance near [Fe/H] = $-$2.0 dex. 
The width of the red giant branch and the location of the candidate Anomalous 
Cepheids on the color-magnitude 
diagram may indicate that the galaxy hosts a complex stellar population with 
stars from 
$\sim$ 13 Gyr to 
as young as $\sim$ 0.6 Gyr.

\end{abstract}

\section{Introduction}
The Canes Venatici I (CVn~I) dwarf spheroidal (dSph) galaxy 
(R.A.= 13$^{\rm h}$ 28$^{\rm m}$ 03.5$^{\rm s}$, decl. = 38$^{\circ}$ 
33$^{\prime}$ 
33.21$^{\prime \prime}$, J2000.0; $\ell$= 74.3$^{\circ}$, b= 79.8$^{\circ}$), 
discovered 
by \citet{zu06}, is one of the new satellite companions to the Milky Way (MW) 
that have been revealed from analysis of 
deep imaging obtained with the Sloan Digital Sky Survey (SDSS) \citep{be07}.  
\citet{zu06} found CVn~I 
to have an absolute magnitude, $M_{v} = -7.9 \pm 0.5$ mag and a central surface 
brightness of $\sim$ 28 mag arc-second$^{-2}$, 
slightly fainter than the well known dSph galaxies 
Draco and UMi, but brighter than other recently discovered dSph galaxies. The 
galaxy has a 
half-light radius $r_h$= $8.5^{\prime} \pm 0.5 ^{\prime}$, with ellipticity 0.38
and overall extent along the major axis of $\sim$ 2 kpc (\citealt{zu06}). 
The SDSS $i, g-i$ color-magnitude 
diagram (CMD) showed CVn~I to be at a large distance (220 kpc), with a 
morphology suggesting a dominant 
old stellar population. However, \citet{dj07},  
applying an automated numerical CMD analysis technique to the SDSS data, also 
suggested 
the possible presence of a younger stellar component about 2.5$-$4 Gyr old.
The CVn~I CMD   
%
exhibits a well-populated horizontal branch (HB), which  
extends across the region of the RR Lyrae instability strip, raising the 
possibility that CVn~I 
contains a significant RR Lyrae population.  RR Lyrae stars, indicative of an 
old stellar population, 
have been found in all the dSph systems identified prior to the recent SDSS 
discoveries, 
as well as in the newly discovered Bo\"{o}tes dwarf \citep{S06,dall06}, which is 
the only one of the new SDSS
dSph's to have been searched thus far for variable stars.
 Although MW globular clusters 
with significant numbers of RR Lyrae stars exhibit the well-established 
Oosterhoff dichotomy \citep{oo39}, 
many (but not all) of the dSph galaxies have Oosterhoff-intermediate properties 
(see \citealt{ca05}).
Notably, the Bo\"{o}tes dSph is in fact an Oosterhoff type II (OoII) system 
\citep{S06,dall06}.


In this {\em Letter} we present the first results of a search for variable stars 
in CVn~I, and use the RR Lyrae stars that 
were discovered to establish its Oosterhoff classification.  We also present a 
$B,V$ CMD of 
CVn~I extending to $V \sim 25$ mag. This is about 3 magnitudes fainter than the 
CMD of \citet{zu06} and
allows us to obtain insight on the stellar components of CVn~I by fitting the 
galaxy CMD 
with theoretical isochrones of different metallicity and age.

\section{Observations and Data Reduction}
$B,V$, and Cousins $I$ time-series photometry of CVn~I was obtained in 2006 May 
and 2007 April and May 
at the 2.3~m Wyoming Infrared Observatory telescope (WIRO), using WIRO-Prime, 
the prime focus CCD 
camera \citep{p02}.  Additional $B,V, I$ photometry was obtained at the 3.5 m 
Telescopio  
Nazionale Galileo (TNG) in 2006 April using the Device Optimized for the LOw 
RESolution  
(DOLORES), and at the 4.2 m William Herschel Telescope in 2007 April using the 
Prime Focus  
Imaging Camera. The WIRO observations cover a field of view measuring 
approximately 17.8 by 17.8 arcmin in size.  
The TNG and WHT observations cover areas of 9.4 by 9.4 arcmin and 16.2 by 16.2 
arcmin, 
respectively. We obtained 43 $V$, 22 $B$ and 15 $I$ frames in total,
corresponding to total exposure times of about 9$^h$, 4.4$^h$ and 2.7$^h$,
in $V$, $B$ and $I$ respectively.
The images were bias subtracted and flat-field corrected using 
IRAF.\footnote[11]{IRAF is 
distributed by the National Optical Astronomical Observatory, which is operated 
by the Association 
of Universities for Research in Astronomy, Inc., under cooperative agreement 
with the National Science Foundation.}.  
Fringing in the $I$-band  images was corrected using a fringe map for each 
image.
Peter Stetson's {\sc Daophot II/Allstar} packages \citep{st87,st92} were used to 
obtain instrumental magnitudes for 
each star.  
These were transformed to the standard system using observations of the Landolt 
standard fields 
PG1323, PG1633, and SA104 \citep{la92} at WIRO, and PG0918 and PG1633 at the 
WHT.  The observed standard 
stars were then used to derive the calibration equations for each telescope.
Typical errors
at the level of the CVn~I HB ($V \sim 22.2$ mag) for the combined
photometry of non-variable stars are
$\sigma_{V}$= 0.01 mag, $\sigma_{B}$= 0.01 mag, and 0.015 mag, 0.03 mag, in the 
WHT and WIRO datasets,
respectively. 

\section{Variable Stars}
We used the $V$-band time series, which had more phase points than the other 
bands, to identify variable 
stars using Peter Stetson's {\sc Allframe/Trial} package \citep{st94}.  A total 
of 84 candidate variables were 
identified.  The majority of these had brightnesses and colors consistent with 
positions on the HB, 
but not all of these stars had a sufficient number of good observations for the 
determination of periods. 
Period searches were carried out using Supersmoother \citep{re94}.  The 
resultant light curves were 
fit to template light curves \citep{ly98} in order to classify the type of 
variable star; these classifications 
were confirmed by eye.  By this procedure we obtained reliable periods and light 
curves for 23 RR Lyrae stars: 18 
fundamental-mode (RRab)  
variables and 5 first overtone (RRc) stars. Comparatively few RRc variables were 
identified compared 
to RRab stars, since we had greater difficulty in establishing reliable periods 
for RRc candidates
due to aliasing problems; their number may therefore be underestimated. 
For these variables there are typically 36 observations in the $V$ band, 
with 22 and 10 observations in the $B$ and $I$ bands, respectively. Three 
brighter variables about 1.5-2 magnitudes
above the galaxy's HB were also identified. They are possibly Anomalous 
Cepheids (ACs), but have incomplete phase coverage. According to the 
position on the galaxy's CMD, the vast majority of the remaining 
58 candidate variables are likely RR Lyrae stars.
Example light curves for some of the identified variables are shown in 
Figure~\ref{f:fig1}. 
The average periods for the RR Lyrae stars are: $\langle P_{ab}\rangle  = 0.60 
\pm 0.01$ days 
($\sigma = 0.02$ days) and $\langle P_{c}\rangle = 0.38 \pm 0.01$ days ($\sigma 
= 0.03$ days).  
The average period for the RRab's suggests that CVn~I is an 
Oosterhoff-intermediate 
object.  The period-amplitude relation for the $V$-band is shown in 
Figure~\ref{f:fig2}. 
The majority of the CVn~I RRab 
stars fall in the region between the Oosterhoff type I (OoI) and Oosterhoff type 
II (OoII) loci, supporting the classification 
of CVn~I as an Oosterhoff-intermediate object. CVn~I would be in this sense like 
the majority of 
dSph systems previously searched for RR Lyrae stars (\citealt{ca04}), but unlike 
the recently 
discovered Bo\"{o}tes system, 
which belongs to the Oosterhoff II class (\citealt{S06,dall06}).
\section{Structure and Color-Magnitude Diagram}
Figure~\ref{f:fig3} shows a 
map of the CVn~I stars in the field of view of the WHT observations. 
A smoothing filter was applied to the data to enhance the stellar densities  
over the background. Although the half-light radius of CVn~I 
($r_h$= $8.5^{\prime} \pm 0.5 ^{\prime}$, \citealt{zu06}) exceeds
the WHT field of view, the bulk of the CVn~I stars appears to be inside 
the elongated structure outlined by the black ellipse of Figure~\ref{f:fig3}, 
whose 
semi-major axis measures $\sim 6.7^{\prime}$.
Figure~4{\it a} shows the $V,B-V$ CMDs of the CVn~I dwarf spheroidal galaxy 
obtained 
from stars in the whole 16.2 $\times$ 16.2 arcmin$^2$ field covered by the 
 WHT observations (Fig. 4{\it a}), and in 3 separate regions corresponding to 
the 
 black ellipse region (Fig. 4{\it b}); the intermediate region between ellipse 
and
 circle (Fig. 4{\it c}); and the region outside the blue circle (Fig. 4{\it d}). 
The 
 three regions cover areas in the ratio 1:1:1.5.
Only stars with
$\sigma_{V}$, $\sigma_{B} \leq 0.10$ mag, $\chi \leq$ 2, and shape-defining 
parameter $\mid SHARP \mid \leq $ 0.5
are plotted in the figure.

The galaxy CMD reaches $V \sim $ 25 mag, is very rich in stars and confirms that 
the dominant 
population in CVn~I is old.
The galaxy has well-populated  
horizontal and red giant branches. The HB stretches across the RR Lyrae 
instability strip, which is entirely
filled by the large number of candidate RR Lyrae stars, and extends 
significantly to the blue. 
The red giant branch (RGB) is a prominent feature of the CVn~I CMD, and exhibits 
some scatter.
Its width suggests 
the existence of a composite population in CVn~I, with stars having some spread 
in metallicity and$/$or age. 
The galaxy CMD is very crowded below $V$ = 24 mag, but nevertheless shows hints 
of a young main 
sequence at $B-V \sim$ 0.2 mag, thus providing support for de Jong et al.'s 
(2007)
earlier suggestion. 
From the average luminosity of the HB ($V_{HB} \sim 22.2$ mag),
the turnoff of the old stellar population is estimated to lie a few tenths below 
the limiting magnitude
reached by our photometry. 
Contamination by field stars is clearly seen in Fig. 4{\it a}, 
with
stars from the MW disk dominating for $B-V >$ 1.2 mag, and stars from the 
Galactic halo 
contaminating the CVn~I CMD at bluer colours.
Since the field of
view covered by our observations does not extend beyond $ \sim 8.9^{\prime}$, we 
do not have a control field devoid of galaxy's stars 
to discuss the field contamination properly. In order to minimize contamination 
by field stars and
to take into account the elongated structure of CVn~I, in Fig. 4{\it b} we have
considered only stars within the black ellipse.
The main features of the galaxy CMD, including 
the blue plume of a young stellar component, shows up much more clearly in Fig. 
4{\it b}. This component becomes increasingly
prominent moving outward (see Figs. 4{\it c} and  4{\it d}). 
The contamination by the MW disk is also significantly reduced in Fig. 4{\it b}. 
Synthetic CMDs of the MW field at the 
position of CVn~I show that a large fraction of the bright stars at $V \sim 20$ 
mag and $B-V \sim 0.4 - 0.5$ mag in  the galaxy CMD 
is accounted for by MW halo stars and foreground galaxies (M. Cignoni, private 
communication).
In  Fig. 4{\it b} we have plotted in red stars of the Galactic globular cluster
M68, from \citet{wa94}, shifted in magnitude (by $-6.58$  mag) and color (by 
$-0.04$ mag) to match the CVn~I 
horizontal and red giant branches. 
The RGB of M68 provides an excellent fit to CVn~I, implying that the old 
population in the galaxy
has a metallicity close to that of M68: ${\rm [Fe/H]}=-2.1$ or $-2.0$ dex on the 
\citet{zw94} 
and \citet{cg97} scales, respectively. 
Since the reddening of M68 is E(B-V)=0.07$\pm$0.01 mag (\citealt{wa94}), the 
$-$0.04 mag color shift
required to match the CVn~I and M68 CMDs implies a reddening 
E(B-V)=0.03$\pm$0.02 mag
for CVn~I. This is slightly larger than the 0.014 $\pm$ 0.026 mag value derived 
for the galaxy from the \citet{sc98} 
maps.

Recent spectroscopic analysis of stars in CVn~I \citep{ib06,ma07} distinguished 
two separate components in CVn~I
with $-2.5 < {\rm [Fe/H]} <  -2.0$ dex and 
$-2.0 < {\rm [Fe/H]} < -1.5$ dex, respectively, on the \citet{cg97} scale.  
These results 
were contested by \citet{sg07}, 
who found no evidence for a second metallicity component 
in CVn~I, based on a sample of stars three times larger 
than that considered by \citet{ib06} and \citet{ma07}.   
Using the \citet{ru97} technique (which is consistent with the 
Carretta \& Gratton 1997 metallicity scale), \citet{sg07} obtained 
${\rm [Fe/H]}=-2.09\pm0.02$ dex, with dispersion $\sigma _{\rm [Fe/H]}$= 0.23 
dex. This metal abundance is 
in good 
agreement with the metallicity we infer from the comparison with M68.
Our CMD and the comparison with M68 do not provide evidence in favor of two 
separate components with
differences in metallicity as large as claimed by \citet{ib06} and \citet{ma07} 
in CVn~I. In order to investigate 
this point further,  
and to possibly disentangle the age and the metallicity effects, we have fitted 
isochrones 
from the PISA 
database\footnote[12]{http://astro.df.unipi.it/SAA/PEL/Z0.html}\citep{car04}
to the CMD of CVn~I
in Fig. 4{\it a}, varying the metallicity in the range from $Z$=0.0002 to 
$Z$=0.0004 and 
the age from 13 to 0.6 Gyr.
Results of the best fit are shown in Figure~\ref{f:fig5}, where we also show the 
variable stars of CVn~I 
in different symbols. 

The best-fit procedure does not favor significant metallicity spreads among the 
CVn~I stars. The galaxy CMD 
is best reproduced by the superposition of 4 subsequent generations of stars 
with roughly the 
same metal
abundance, $Z$=0.0002 ([Fe/H]=$-$2.0 dex). They include:  a 13 Gyr component 
(red isochrone) accounting for the redder RGB, the HB and the RR Lyrae stars; a 
5 Gyr population (blue isochrone) 
producing the bluer RGB and the fainter portion of the red clump; 1.5 Gyr stars 
(green isochrone) providing part of the
blue plume and the brighter portion of the red clump; and, finally, a 0.6 Gyr 
component 
(black isochrone) producing the bluest part of the blue plume and the ACs. A 
deeper CMD is needed to
fully confirm this interpretation. 

\section{An Improved Distance Estimate for CVn~I}
The intensity-weighted mean magnitude of the Canes Venatici RR Lyrae stars is 
${\rm \langle V\rangle}$ = 22.17 $\pm 0.02$ mag 
(with a dispersion of 0.07 mag among the 23 stars). We adopt an absolute 
magnitude of 
$\rm{M_{V}}=0.46\pm0.03$ mag for RR Lyrae stars with metallicity 
$\rm{[Fe/H]}=-2.1$ dex \citep{cc03}.  We assume
a reddening for CVn~I of $\rm{E(B-V)}=0.014 \pm 0.026$ mag \citep{sc98}, with 
consequent absorption in the 
$V$ band of 
about 0.04 mag.  We thus find a distance modulus of $\mu_{0}=21.67\pm0.06$ mag, 
which corresponds to a 
distance of $d=216^{+7}_{-5}~\rm{kpc}$.  This agrees with the distance of 
$220^{+25}_{-16}~\rm{kpc}$ 
found in \citet{zu06}, which is also based on the \citet{sc98} reddening. 
If we adopt instead the value of $E(B-V)$=0.03 $\pm$ 0.02 mag inferred from the 
comparison with M68, 
we obtain a distance modulus of $\mu_{0}=21.62\pm 0.06$ mag, and in turn a 
distance of 
$d=210^{+7}_{-5}~\rm{kpc}$.
%
%

\section{Conclusions}
We have identified and obtained periods and light curves for 18 RRab stars and 5 
RRc stars in the Canes 
Venatici I dSph galaxy. 
We also identified three potential Anomalous Cepheids and 58 additional 
candidate variable stars that according to their 
position on the galaxy CMD are very likely RR Lyrae stars.  The 
average period of the CVn~I RRab stars for which we have complete and reliable 
light curves and their location on the period-amplitude 
diagram suggest that CVn~I is an Oosterhoff-intermediate type object.  
Thus CVn~I seems to follow the trend of the other ``classic'' dSphs 
\citep{ca04}. This similarity is strengthened in that CVn I seems to contain a complex stellar population with components of different age in the range from 13 to 0.6 Gyr.

\bigskip
We thank Michael Pierce for assistance with the WIRO Prime camera.  HAS thanks 
the Center for Cosmic Evolution 
and the U.S. National Science Foundation for support under grant AST0607249.  
M.C. is supported by Proyecto 
Fondecyt Regular \#1071002.  CTR was funded by Wyoming NASA Space Grant 
Consortium, NASA Grant \#NNG05G165H.  
T.~C.~B. acknowledges support by the US National Science Foundation under grants 
AST 06-07154 and AST 07-07776, 
as well as from grant PHY 02-16783; Physics Frontier Center/Joint Institute for 
Nuclear Astrophysics (JINA).  
G.C. acknowledges support by PRIN-INAF 2006 (PI G. Clementini).  K.K. 
acknowledges support from National 
Science Foundation grant AST-0307778.

\clearpage

\realfigure{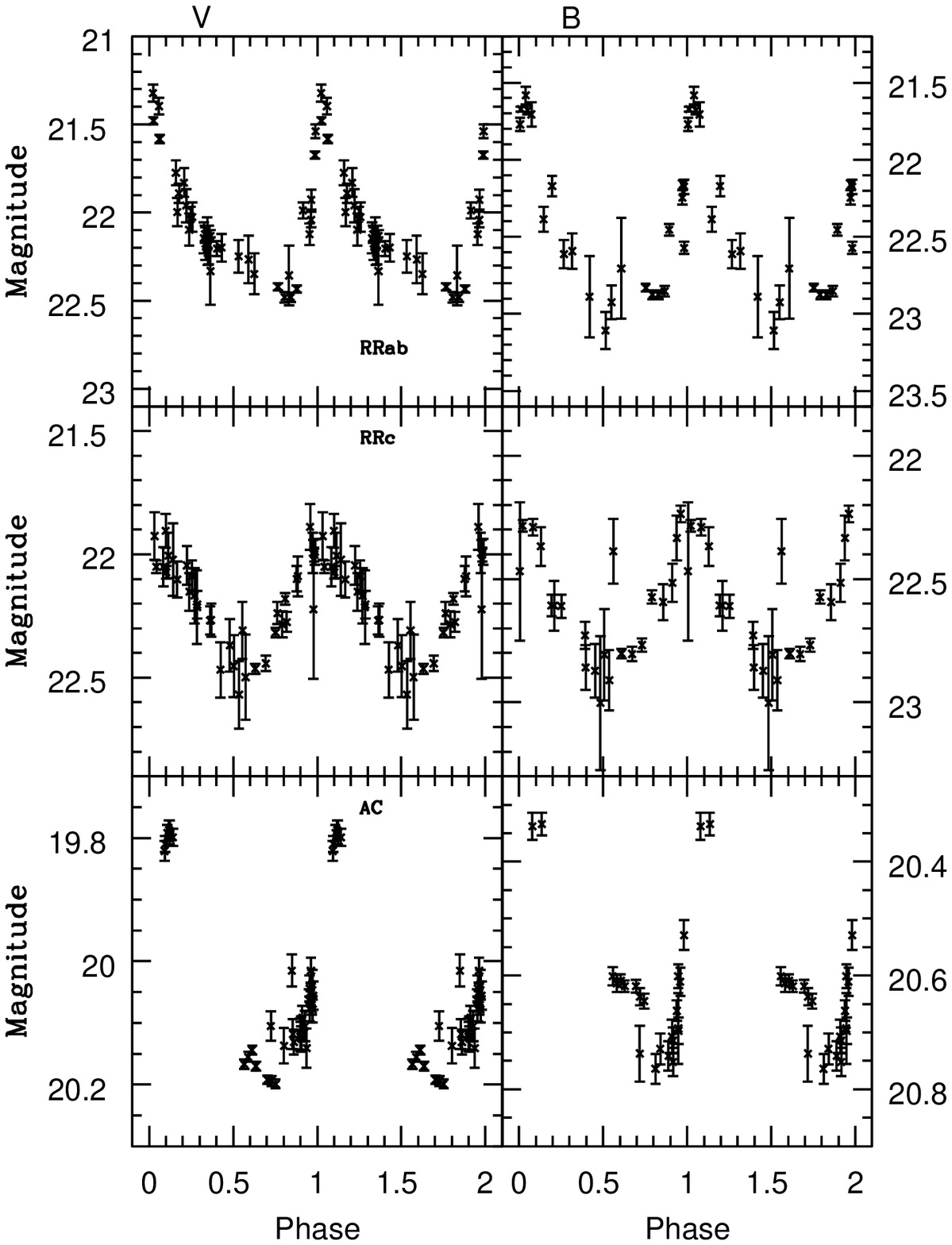}
{$V$ and $B$ light curves of variable stars in CVn~I.
$Top:$ RRab star with $P$=0.63 days.
$Middle:$ RRc star with $P$=0.40 days.  $Bottom:$ candidate Anomalous Cepheid 
with a possible 
period of 1 day.}{f:fig1} 

\clearpage

\realfigure{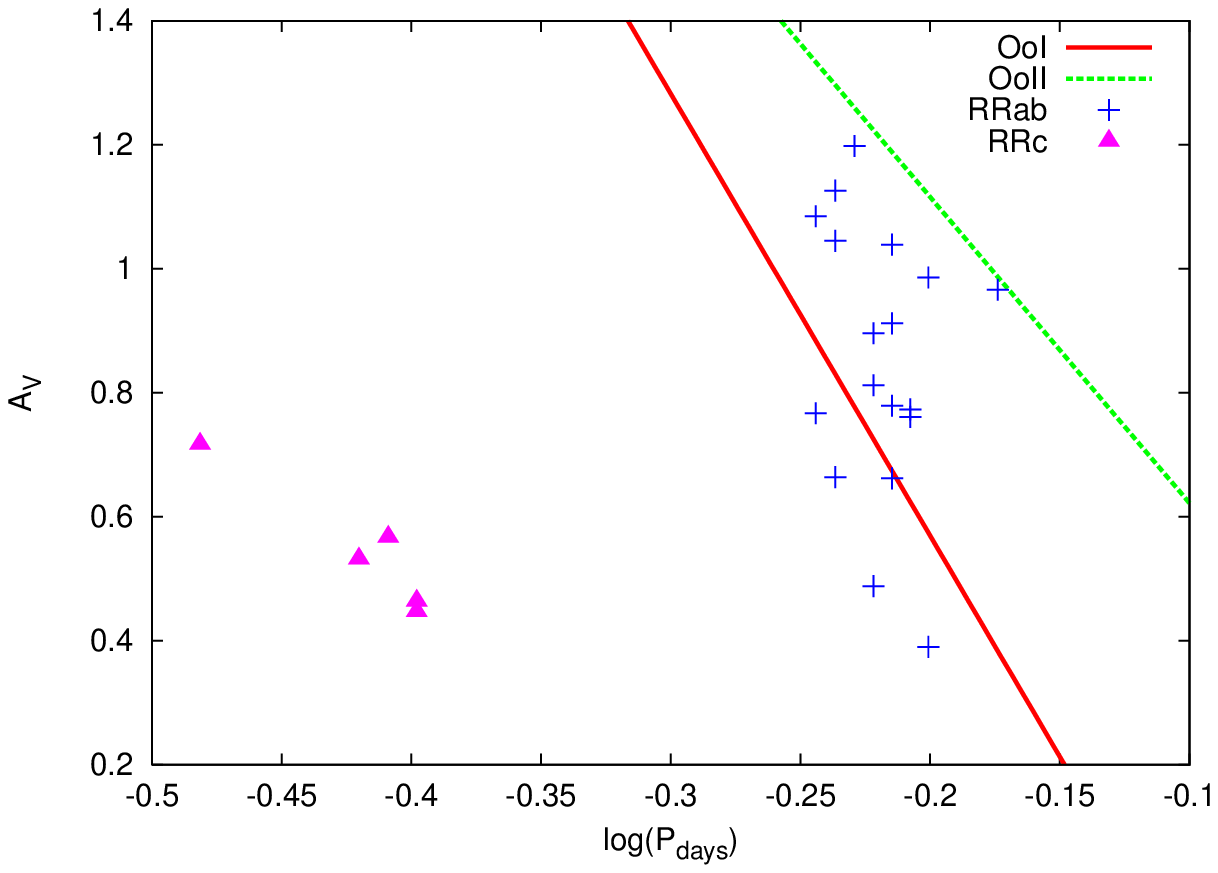}
{Period-amplitude diagram in the $V$ band for the CVn~I RR Lyrae stars. 
RRab stars are indicated by blue crosses and RRc stars are indicated by purple 
triangles.  The solid
and dashed lines are the positions of Oosterhoff type I (OoI) and Oosterhoff 
type II
(OoII) Galactic globular clusters from \citet{cr00}. 
}{f:fig2}

\clearpage

\realfigure{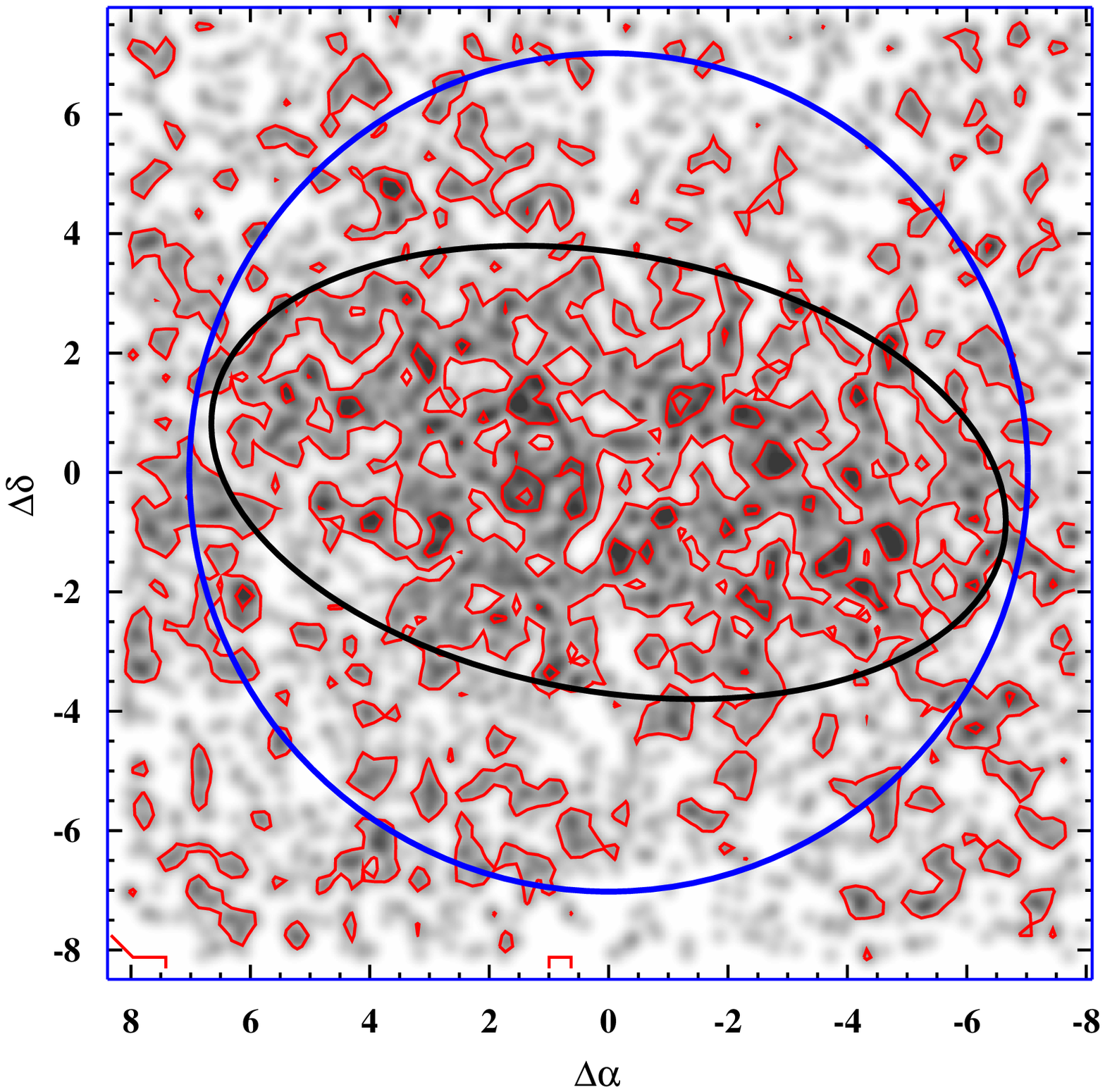} 
{Smoothed 
map of the galaxy stars in the field of view of the WHT observations, in 
differential 
R.A. and declination from the CVn~I dSph center. The bulk of the
CVn~I stars is within the black ellipse which has semi-major axis $r$= 6.7 
arcmin and ellipticity $e$=0.55. The blue
circle has a radius of $r$ = 7 arcmin. 
 }{f:fig3}

\clearpage

\begin{figure*}
\includegraphics[scale=.90]{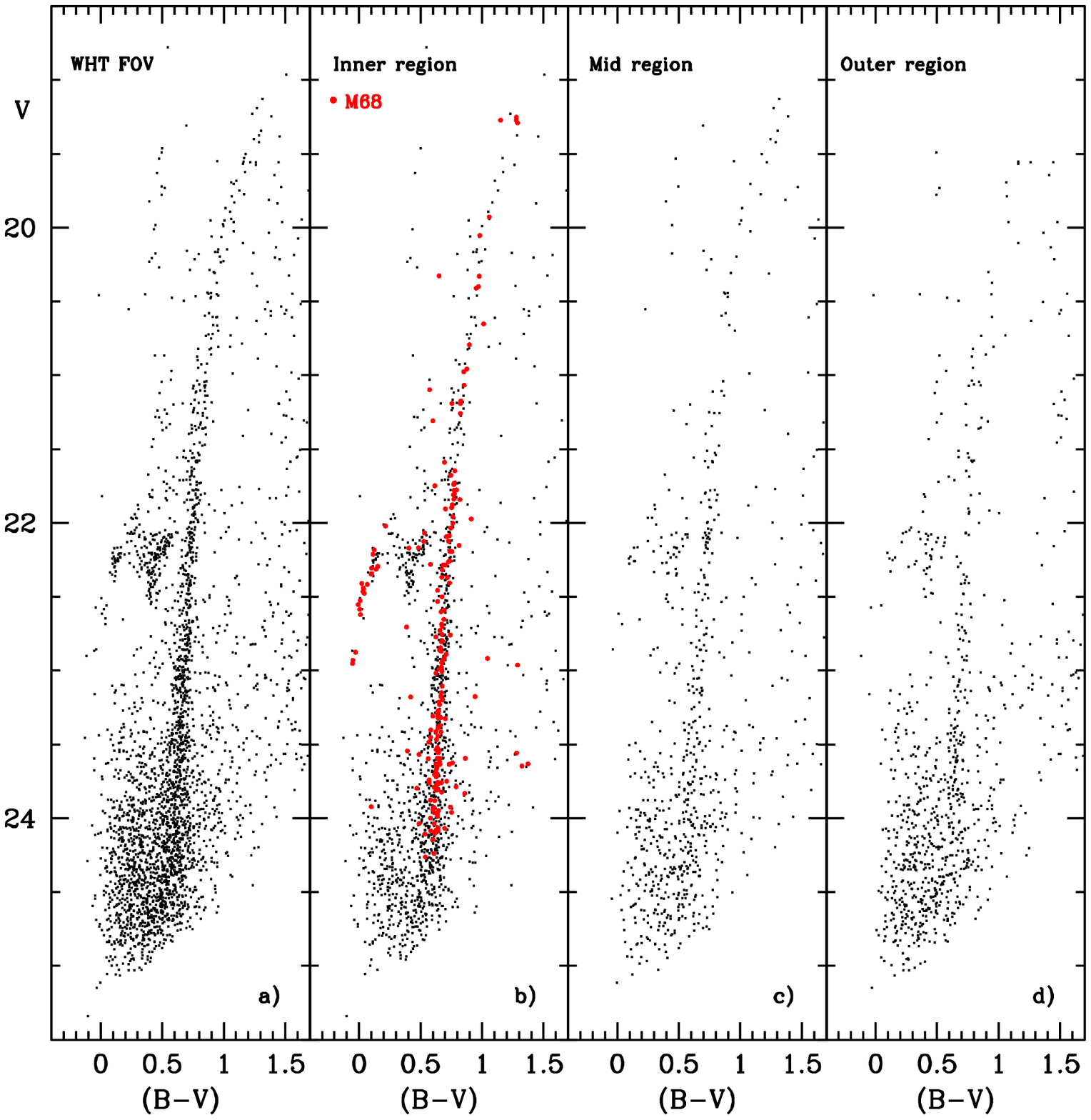}
\caption{$V,B-V$ CMDs of the Canes Venatici I dwarf spheroidal galaxy obtained 
from stars
in the whole 16.2 $\times$ 16.2 arcmin$^2$ field covered by the 
 WHT observations (Fig. 4{\it a});  and in 3 separate regions 
 corresponding to the 
 black ellipse (Fig. 4{\it b}); the intermediate region between ellipse and
 blue circle (Fig. 4{\it c}); and the region outside the blue circle (Fig. 4{\it 
d}) of Figure~\ref{f:fig3} .
Plotted in red are stars of the Galactic globular cluster
M68, after \citet{wa94}.
}
\label{f:fig4}
\end{figure*}

\clearpage

\realfigure{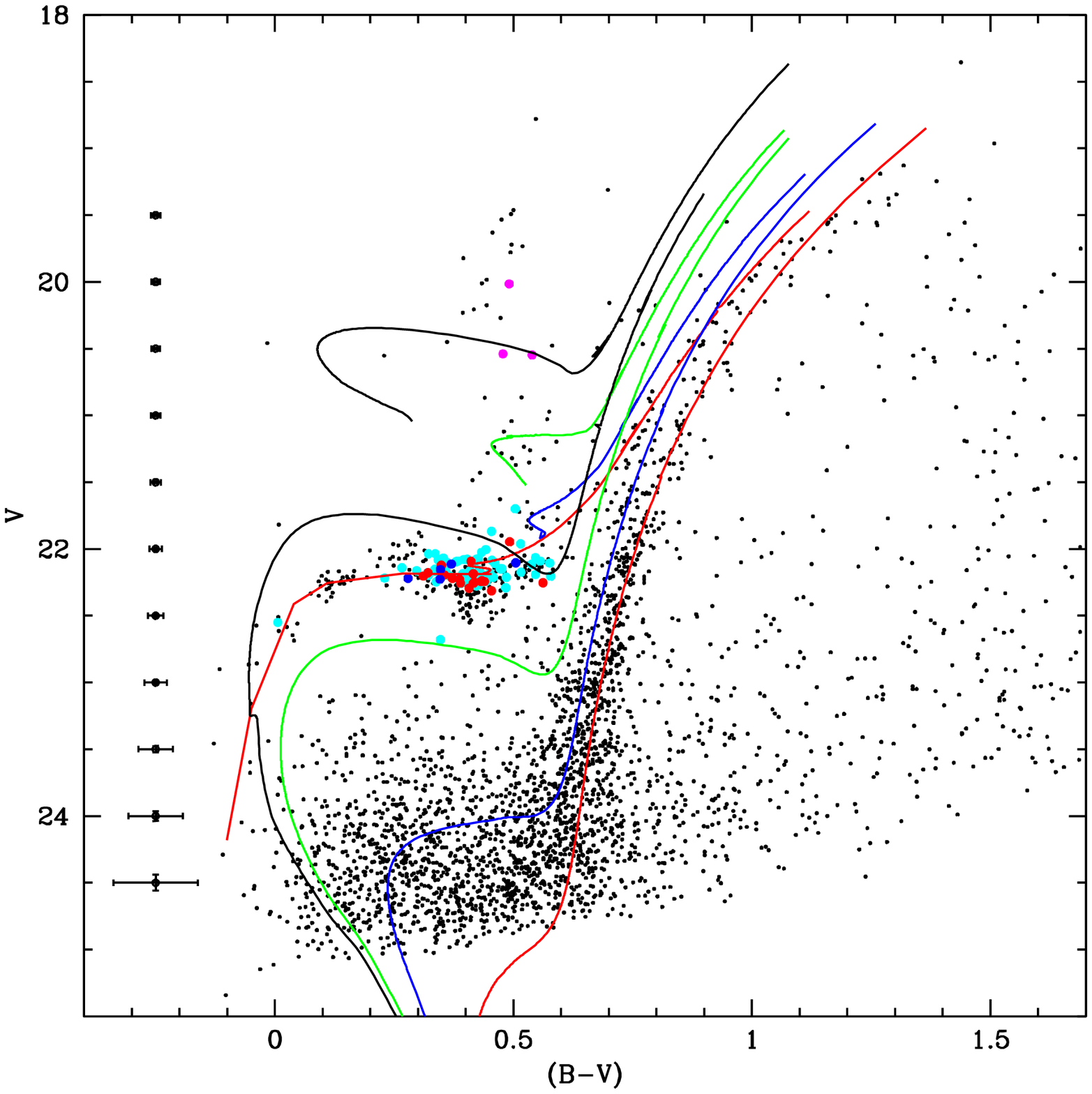}
{Best fit of the CVn~I CMD in Fig. 4{\it a}, using isochrones from the Pisa 
dataset 
with $Z$=0.0002 ([Fe/H]=$-$2.0 dex),
and 4 age components; 13 Gyr ({\em red line}), 5 Gyr ({\em blue line}), 1.5 Gyr 
({\em green line}), and
0.6 Gyr ({\em black line}). The CVn~I variable stars are plotted with different 
symbols: red, RRab stars; blue, RRc stars;
cyan, candidate variables for which we were not able to derive reliable periods; 
purple, candidate ACs.
Typical error bars of the photometry are provided on the left-hand 
side.}{f:fig5}

%
%
%
%
%
%

\end{document}